
........................................................................

   This paper can be printed only with special macros. The figures
are not attached. If the text below is not enough, simply request a
hardcopy from the authors.

\input jnl.tex
\input reforder.tex
\input eqnorder

\def\jt{\rm J/t}

\def\quarter{\langle n \rangle = 1/2}
\def\chid{\chi^d_{sup}}
\def\chis{\chi^s_{sup}}
\def\dx2y2{\rm d_{x^2-y^2}}
\def\ds{\rm D_s}
\def\drude{\rm D_{Drude}}

\null
\vskip -.75in
\vskip .5in
{\singlespace
\smallskip
\rightline{ October, 1992}
}

\vskip 0.1in

\title INDICATIONS OF ${\bf {\rm d_{x^2 - y^2}}}$ SUPERCONDUCTIVITY
       IN THE TWO DIMENSIONAL t-J MODEL

\vskip 0.2in

\author  E.~DAGOTTO and J.~RIERA$^\ast$

\affil Department of Physics,
       Center for Materials Research and Technology and
       Supercomputer Computations Research Institute,
       Florida State University,
       Tallahassee, FL 32306

\vskip 0.1in

\abstract { Superconducting correlations in
the two dimensional ${\rm t-J}$ model at zero
temperature are evaluated using numerical techniques.
At the fermionic density $\langle n \rangle \sim 1/2$,
strong signals of $\dx2y2$ superconductivity were observed in
the ground state. These conclusions are based on
a study of
static pairing correlations, the Meissner effect,
flux quantization, and other indicators of
superconductivity.
It is argued that these results can be explained using
a spin dimer ``liquid'' state.
A phase diagram of the two dimensional ${\rm t-J}$
model is presented.
}

\vskip 0.4truecm

\line{PACS Indices: 75.10.Jm, 75.40.Mg, 74.20.-z\hfill }

\vskip 0.4truecm

{\singlespace
\noindent
$\ast$) Permanent address:
Departamento de F\'isica, Facultad de Ciencias Exactas,
Av. Pellegrini 250, 2000 Rosario, Argentina.}

\endtitlepage

The study of high-${\rm T_c}$ superconductors continues attracting
considerable attention. Recent calculations suggest that
non s-wave symmetry pairing interactions with nodes, may explain some of the
unusual properties of the cuprate compounds, in particular the
behavior of relaxation rates in the ${\rm Y Ba_2 Cu_3 O_7}$
material, as well as the systematic presence of spectral
weight inside the superconducting gap.\refto{exper}
More specifically, the possibility
of $\dx2y2$ superconductivity in the cuprate materials
has been recently discussed.\refto{pines,dwave}
Most of these calculations have been performed without specifying the
details of
the interaction, but analyzing a BCS-like gap equation for
different pairing symmetries. Thus, it would be important to find a
realistic model of strongly interacting electrons having a $\dx2y2$
symmetric
superconducting state as ground state. From the properties
of this state, dynamical responses of a d-wave condensate could be
studied, and concrete predictions would be made to contrast theory with
experiments.
In this scenario,
numerical studies are important to decide
whether a given electronic model presents a superconducting phase,
specially since the strongly
interacting character of several realistic models makes most
analytical approximations questionable. The
one band Hubbard model is a typical example of these problems, i.e.
while for some time it was assumed that the ground state at finite
hole doping superconducts, Quantum
Monte Carlo simulations have not supported these claims.
Then, the issue of whether purely electronic models of
high-${\rm T_c}$ materials present a superconducting ground
state is still open.

The purpose of this paper is to present numerical results
suggesting that the widely studied
two dimensional ${\rm t-J}$ model has a superconducting phase
in a previously unexplored
region of parameter space. The symmetry of the
condensate is $\dx2y2$, and thus this model may become
a physical realization
of the d-wave pairing scenarios recently proposed in the
literature.\refto{dwave,pines} The superconducting
phase observed here appears near the well-known region of
phase separation of the ${\rm t-J}$ model.\refto{emery,putikka}
The possible presence of superconducting correlations near phase
separation has been recently discussed in other contexts
and theories,\refto{super,kivelson,adriana} but
here the first numerical indications are provided that this phenomenon
may occur in the ground state of a realistic model of strongly correlated
electrons.
The ${\rm t-J}$ model is defined by the Hamiltonian,
$$
{\rm H =
{\rm J } \sum_{{\bf \langle i j\rangle }}
( {{\bf S}_{\bf i}}.{{\bf S}_{\bf j}} - {1\over4} n_{\bf i} n_{\bf j} )
- {\rm t} \sum_{{\bf \langle i j \rangle},s}
({\bar c}^{\dagger}_{{\bf i},s} {\bar c}_{{\bf j},s} + h.c.) },
\tag 1
$$
\noindent where
${\rm  {\bar c}^{\dagger}_{{\bf i},s}}$
denote $hole$
operators;
${\rm n_{\bf i} = n_{{\bf i},\uparrow} + n_{{\bf i},\downarrow} }$;
and square clusters of ${\rm N}$ sites with
periodic boundary conditions are considered.
The rest of the notation is standard. Here,
efforts have been concentrated on the exact diagonalization of
$4 \times 4$ lattices, although preliminary results for
clusters of 20 and 24 sites are available. It has been
repeatedly shown in the literature that these cluster sizes are
large enough to capture the essential qualitative physics of
several models of strongly correlated electrons. Besides, no
other available (and unbiased)
numerical technique can handle the involved
calculations that have been carried out for the
${\rm t-J}$ model without making assumptions about
the properties of the ground state.
To search for indications of superconductivity,
let us define the singlet pairing operator
$\Delta_{\bf i} = {c}_{{\bf i},\uparrow}
(  {c}_{{\bf i+{\hat x}},\downarrow} +
   {c}_{{\bf i-{\hat x}},\downarrow} \pm
   {c}_{{\bf i+{\hat y}},\downarrow} \pm
   {c}_{{\bf i-{\hat y}},\downarrow} )$,
where $+$ and $-$ corresponds to
extended-s and $\dx2y2$ waves, respectively,
and ${\bf {\rm {\hat x},{\hat y}}}$ are unit vectors along the axis.
The pairing-pairing correlation function
${\rm C({\bf m}) = {1\over N} \sum_{{\bf i}} \langle \Delta^{\dagger}_{\bf i}
\Delta_{{\bf i + m}} \rangle}$,
and its susceptibility $\chi^\alpha_{sup}
= {\rm \sum_{{\bf m}} C({\bf m})}$ have been calculated
(where $\alpha = d$ corresponds to
$\dx2y2$ wave, and $\alpha = s$ to extended-s wave).
$\langle \rangle$ denote expectation values in the ground state, which
is obtained using the Lanczos method.

$\chi^d_{sup}$ is shown in Fig.1a as a function of $\jt$, for
several densities. In this study, it was observed that the $\dx2y2$ wave
susceptibility dominates, presenting at $\quarter$
a sharp peak at $\jt \sim 3$. By
analyzing several spin and hole correlations, and following other
criteria\refto{assa} it was
verified that the fast decay of $\chid$ after the peak is induced by
the transition to the phase separated region.
Changing the fermionic density,  it was observed that
$\chid$ has its maximum value at $\quarter$, as shown in Fig.1a.
$\chis$ has been also evaluated
at $\quarter$. This susceptibility peaks
at approximately the same position as $\chi^d_{sup}$ does, but with
a smaller intensity.\refto{future}
The pairing-pairing correlations as a function of distance are
shown explicitly in Fig.1b in the region where the
susceptibilities have a sharp maximum, i.e $\jt = 3.0$ and
$\quarter$. As expected from the behavior of $\chi^d_{sup}$,
Fig.1a, the
dominant correlation functions at the maximum distance on
the $4 \times 4$ cluster corresponds to $\dx2y2$ symmetry.
On the other hand, the correlations
for extended-s operators are strong at short distances, but decay
rapidly at large distances.
For completeness, in Fig.1b the
correlation corresponding to ${\rm d_{xy}}$ symmetry is
also presented.\refto{dxydxy}
These correlations seem more heavily suppressed than for
the $\dx2y2$ and extended-s channels. In Fig.1c, the d-wave
pairing correlations are shown at $\quarter$, as a function of
$\jt$. Their maximum value is obtained at the same coupling where
$\chid$ peaks, as expected.

The symmetry of the ground state (obtained with the Lanczos
method) under a rotation of the lattice
in $\pi/2$ has been studied.
In the region where
superconducting correlations exists, the ground state is $odd$
under this operation,
but it is invariant under reflexions
with respect to the x and y-axis. Then, the ground state at $\quarter$
belongs to the $B_{1g}$ representation of the $C_{4v}$ group,
usually denoted by $\dx2y2$, in agreement with the previous conclusions
studying pairing correlations. However, it is worth noticing
that the
state with the lowest energy in the subspace invariant under
rotations and reflexions
(i.e. s-wave) is close in energy to the ground state. More
specifically, at
$\quarter$ and $\jt=3.0$, the energy of the d-wave ground state
is $-26.413 {\rm t}$, the lowest s-wave state has energy
$-26.127 {\rm t}$, while, for comparison,
the lowest spin one state in the spectrum carries an
energy $-25.188 {\rm t}$. Then, a scenario
where ${\rm s+id}$ pairing occurs in the bulk limit
is not excluded, although it is
clear that the $\dx2y2$ correlations seem stronger.

The pairing correlations found in the
two dimensional ${\rm t-J}$
model suggest the existence of a superconducting
phase near phase separation. To complete the
analysis, it is necessary to show that a $Meissner$ effect
occurs in that region.
Recent progress\refto{scalapino} in
the analysis of the superfluid density, $\ds$, using linear
response theory allow us to carry out such a study using
techniques similar to those required to analyze the
Drude peak in the optical conductivity, $\sigma (\omega)$,
of strongly interacting
electrons.\refto{sigma}
Following Scalapino et al.,\refto{scalapino}
it can be shown that $\ds$ is given by
$$
{\rm {{\ds}\over{2 \pi e^2 }} =
{{\langle -T \rangle} \over{4N}}  - {{1}\over{N}}
\sum_{n \neq 0} {1\over{E_n - E_0}}
|\langle n | j_x({\bf q}) |0\rangle |^2 },
\tag 2
$$
\noindent where ${\rm e}$ is the electric charge;
the current operator in the x-direction with momentum ${\rm {\bf q}}$
is given by ${\rm j_x ({\bf q})} = \sum_{{\bf l},\sigma}
e^{i {\bf q}.{\bf l}}
( {\bar c}^{\dagger}_{{\bf l},{\sigma}}
  {\bar c}_{{\bf l + {\hat x}},{\sigma}} -
  {\bar c}^{\dagger}_{{\bf l}+{\bf {\hat x}},{\sigma}}
  {\bar c}_{{\bf l},{\sigma}} )$; $\langle {\rm - T} \rangle$ is the
kinetic energy operator of Eq.(1); ${\rm | n \rangle}$ are eigenstates
of the ${\rm t-J}$
Hamiltonian with energy ${\rm E_n}$ (where ${\rm n = 0}$ corresponds
to the ground state), and the rest of the notation is standard.
The momentum ${\rm {\bf q }=(q_x,q_y)}$ of the current operator
is selected such that
${\rm q_x}=0$ and ${\rm q_y \rightarrow 0}$.
The constraint of having
an infinitesimal but nonzero ${\rm q_y}$ is necessary to
avoid a trivial cancellation of $\ds$
due to rotational and gauge
invariance.\refto{scalapino}
On the $4 \times 4$ cluster, the minimum value of ${\rm q_y}$
is $\pi/2$, and that is the momentum used in the present
analysis. $\ds$ given by Eq.(2) can be
evaluated numerically using the continued fraction expansion technique
previously used to extract dynamical information from
finite clusters.\refto{sigma}
In Fig.2a, $\ds$ is shown as a function of $\jt$ for
several densities. In good agreement with $\chi^d_{sup}$,
the superfluid density $\ds$ presents a sharp maximum
in the neighborhood of phase separation at $\quarter$
giving support to the
previous conclusions regarding the existence of superconductivity in
this model.\refto{test}
It is interesting to note that the signal is stronger
for lower densities, e.g. $\langle n \rangle = 0.25$, perhaps due to
the higher mobility of pairs in that regime. In the phase
separated region, $\ds$ is small, as expected.

The resistivity of the model has also been analyzed.
{}From previous studies of the optical conductivity in strongly
interacting models,\refto{sigma} it can be shown that
the Drude peak
is given by a simple modification of Eq.(2), i.e. it is enough to
replace $\ds$ $\rightarrow$ $\drude$, and consider zero momentum,
${\bf q}=(0,0)$, in the current.\refto{sigma,scalapino}
In Fig.2b, $\drude$ is shown as
a function of $\jt$, for several densities. In the region of
phase separation, the conductivity is small as expected, while
for smaller values of $\jt$, the Drude peak is considerably larger.
A finite value of $\drude$ in the bulk limit implies a zero
resistivity, $\rho = 0$, since
$\sigma (\omega \rightarrow 0) =
{\rm \rho^{-1} = D_{Drude}} \delta (\omega)$.
Fig.2b suggests that this result will
hold not only in the superconducting region, but it will survive a
further reduction of the coupling into the small $\jt$ regime, i.e.
even in a phase without pairing.
This example shows that $\rho$ is not enough to
distinguish between a ``perfect metal'' and a ``superconductor'',
and thus the previously discussed study of the superfluid density is crucial
to show unambiguously the presence of superconducting
correlations in the model.\refto{scalapino}
To further complete the present analysis, the response of the system
to an external magnetic $flux$ $\phi$ was studied. For this
purpose, a phase factor $e^{i \phi /N}$ is introduced
in the kinetic energy hopping terms of Eq.(1), but only
in the x-direction. This is equivalent to allowing a nonzero
flux across one of the ``holes'' of the torus.\refto{didier}
In Fig.3a, the ground state energy $\Delta E(\phi)=
E(\phi) - E(\phi = 0)$, in the zero momentum subspace,
is shown as a function of $\phi$, at density $\quarter$.
In the region of pairing, $\jt = 3.0$, the energy presents
two minima, one located at $\phi = 0$ $( mod$ $2\pi)$,
and a nontrivial one at $\phi = \pi$, signaling
the presence of carriers with charge $2e$
in the ground state, in agreement with the analysis based on the
pairing correlations.

What is the nature of the superconducting state at
$\langle n \rangle \sim 1/2$?
It is reasonable to expect
that the same force that produces phase separation, is
responsible for superconductivity. Actually, if
two electrons are considered on an otherwise empty lattice, they form a
bound state at ${\rm J/t =2}$, and at low electronic density this same
attraction leads to phase
separation when the coupling is increased.\refto{emery}
In this respect,
the antiferromagnetic coupling should be considered as an $attractive$
interaction in the Hamiltonian at low densities, and thus the
presence of superconductivity in the model is easily understood.
At large $\jt$, pairs of electrons (at least
for small $\langle n \rangle$) are expected
to have a size comparable to the range of the force,
namely approximately
one lattice spacing, and in this respect the superconducting correlations
discussed in this paper may be
of the bipolaronic type. The pairs
should be coupled in short spin singlets forming $dimers$.
To check these ideas let us analyze
the spin-spin correlations. In this scenario, each electron is
coupled with only one other particle in a spin singlet, but
due to rotational invariance, that particle can be
located at any
of the four possible nearest neighbors. Then, the correlation
at distance of one lattice spacing,
should be $1/4$ of the on-site correlation,
and it should vanish at larger distances. The results shown in Fig.3b
obtained at $\quarter$ and $\jt = 3.0$,
are in excellent agreement with this picture. Another issue to
address is the possible formation of a ``crystal'' structure.
Is there any special
order in the position of these dimers? For that purpose
hole-hole correlations,
${\rm h(m) = \langle n_h(0) n_h(m) \rangle }$ (where ${\rm n_h}$ is the
hole number operator) were studied,
i.e. once a hole is located at a given
site $0$, then correlations with other holes are evaluated.
Asymptotically, ${\rm h(m)}$
should decay to $\langle n \rangle$ at large distance.
In Fig.3c, ${\rm h(m)}$ is shown
at $\quarter$ and $\jt = 3.0$. The hole-hole
correlations rapidly decay to its asymptotic value, showing that
there is no special pattern in the hole distribution (or, equivalently
at $\quarter$, in the electronic distribution).
The spin-gap in the neighborhood of phase
separation has also been studied.
Although to obtain conclusive results a careful
finite size scaling analysis is necessary,
the data suggest the presence of
a finite spin-gap,\refto{jap} compatible with the idea of small size
dimers as responsible for the superconducting correlations.\refto{gold}
To summarize these ideas, in Fig.4a a snapshot of
the ``dimer liquid'' state that here is claimed to be
compatible with the numerical
results is presented.\refto{rvb}
Note that this state explains the observed
maximum that $\chi^d_{sup}$ presents at $\quarter$,
since at that density the model has
the maximum number possible of mobile dimers.
At smaller densities, there are fewer dimers contributing to the
signal; while closer
to $\langle n \rangle = 1$, the pairs of electrons have
less mobility for lack of space.
A similar state, although defined in a rigid crystal pattern,
was previously discussed in the context of the ${\rm t-J-V}$
model.\refto{kivelson,super}

Summarizing, in this paper
numerical evidence suggesting
that the ${\rm t-J}$ model in two dimensions
has a superconducting phase at zero temperature has
been discussed. The
pairing correlations are the strongest at density
$\quarter$, and near phase separation.\refto{emery,putikka}
The symmetry of
the pairing state corresponds to $\dx2y2$. The Meissner
effect, as well as flux quantization calculations
support this scenario. The size of the pairs seem
small at $\quarter$ as suggested by the spin-spin
correlations, and they are in a disordered state. Thus,
a liquid of dimers may represent the physics of this
condensate.
These results have several implications:
i) they are the first numerical
evidence that the ${\rm t-J}$ model
superconducts in two dimensions. Previous numerical
studies\refto{sigma} concentrated their efforts near $\langle n \rangle \sim
1$,
but in that regime the signal for superconductivity
would be too weak to be detectable;
ii) In addition, it was found
that the symmetry of the superconducting condensate
is $\dx2y2$, and thus this model may become a
realization of recent proposals to explain the
phenomenology of high-${\rm T_c}$ materials
making use of non s-wave pairing interactions with
nodes.\refto{pines,dwave} Based on the present
calculation and others\refto{emery,putikka}
the currently available information for the phase diagram of the
two dimensional ${\rm t-J}$ model at zero temperature
is sketched in Fig.4b. The notation is explained in
the caption. The ``binding'' region denotes a regime where pairs
are formed, but they are not condensed in a superconducting state.
The details
of this phase diagram close to half-filling are
more difficult to
address numerically than at $\quarter$. However,
the possibility that the model superconducts also at low hole
doping is not excluded. Whether there is an analytical continuation
between $\quarter$ and large $\jt$, and densities closer to half-filling
and smaller couplings is a crucial issue for the success of the
${\rm t-J}$ model as a phenomenological model of high-${\rm T_c}$
superconductors. This important subject will be addressed in future
publications.

This work benefited from
useful conversations with A. Moreo, M. Luchini, D. Scalapino, R. Laughlin,
and F. Ortolani.
The authors thank the
Supercomputer Computations Research Institute (SCRI) and NCSA,
Urbana, Illinois, for their support.

\endpage

\references

\refis{gold} In addition,
studying the ground state in the subspace with two less
electrons, clear indications of a ``Goldstone particle'' in the
spectrum with the proper symmetry for d-wave pairing were observed,
due to the presence of superconductivity in the
ground state (see A. Moreo, Ref.\cite{adriana}).

\refis{dxydxy}
A possible $\dx2y2$ + i ${\rm d_{xy} }$ pairing
mechanism in the
${\rm t-J}$ model has been recently suggested by R. Laughlin
(private communication).
For ${\rm d_{xy}}$ symmetry, the pairing
operator is
odd under both axis-reflexions and $\pi/2$ rotations, and
the carriers are located at a distance of $\sqrt{2}$ lattice
spacings from each other.

\refis{future} Details will be presented in an extended version of this
paper, in preparation.

\refis{test} A good test of the programs used to compute $\ds$ consists
of taking ${\bf q} = (\pi/2,0)$ as the current momentum. This is
equivalent
to using a ``trivial'' gauge field, i.e. one
that does not induce neither electric nor
magnetic fields in the problem. In such a case $\ds$ should
vanish identically (we thank D. Scalapino for this remark).

\refis{rvb} Note the similarity with the RVB states discussed, for
example, by  L. Pauling, Proc. R. Soc. London {\bf A 196}, 343 (1949);
P. W. Anderson, Science {\bf 235}, 1196 (1987); and S. Kivelson, D.
Rokhshar, and J. Sethna, Phys. Rev. {\bf B 35}, 8865 (1987).

\refis{didier} For more details
see D. Poilblanc et al., \prb 44, 466, 1991.

\refis{sigma} {E. Dagotto et al., \prb 45, 10741, 1992; and
references therein.}

\refis{scalapino} D. Scalapino, S. White, and S. Zhang, \prl 68, 2830,
1992.

\refis{super} E. Dagotto and J. Riera, FSU preprint, to appear in PRB.

\refis{jap} Even if the spin-gap vanishes for the ${\rm t-J}$ model, the
addition of a small density-density repulsion is enough to open a gap.
See Ref.\cite{super}, and
H. Tsunetsugu and M. Troyer, private communication, for similar
conclusions in the 1D ${\rm t-J-V}$ model.

\refis{adriana} A recent study of the spectrum of the ${\rm t-J}$ model
suggested that superconductivity may exist near phase separation in 2D
(A. Moreo, \prb 45, 4907, 1992).
Also, in the context of a three band Hubbard
model analyzed with slave bosons and large-N techniques,
superconductivity
near phase separation was observed by M. Grilli, R. Raimondi, C.
Castellani, C. Di Castro and G. Kotliar, \prl 67, 259, 1991. This
instability was found in the $A_{1g}$ and $B_{2g}$ channels i.e.
not in the $\dx2y2$ channel ($B_{1g}$). In 1D problems,
superconducting correlations are dominant near phase separation in
the ${\rm t-J}$ model. See for example
M. Ogata, et al., \prl 66,
2388, 1991; F. Assaad and D. W\"urtz, \prb 44, 2681, 1991;
M. Imada, J. Phys. Soc. Jpn. {\bf 58},
3752 (1989); A. Sudbo et al., preprint.
Of course, in one dimension, no
long-range order can exist, and there is no Meissner effect.

\refis{assa} See F. Assaad and D. W\"urtz, see Ref.\cite{adriana}, where
correlation functions at a fixed density are used, instead of
energies at several densities as in the compressibility.


\refis{kivelson} S. Kivelson, V. Emery, and H. Q. Lin, \prb 42, 6523,
1990, discussed the possibility of s-wave superconductivity in the
${\rm t-J-V}$ model, at large ${\rm V/t}$.

\refis{emery} V. Emery, S. Kivelson and H. Q. Lin, \prl 64, 475, 1990.

\refis{putikka} W. Putikka, M. Luchini and T. M. Rice, \prl 68, 538, 1992.

\refis{exper} E. T. Heyen, et al., \prb 43, 12958, 1992; T. Pham et al.,
\prb 44, 5377, 1991; J. M. Valles Jr., et al., \prb 44, 11986, 1991.

\refis{dwave} N. Bulut and D. Scalapino, \prl 67, 2898, 1991; and
references therein.

\refis{pines} P. Monthoux, A. Balatsky, and D. Pines, \prl 67, 3448,
1991.

\endreferences

\endpage

\vskip 1cm

\bigskip
\centerline{\bf Figure Captions}
\medskip

\item{1a}  $\dx2y2$
           superconducting susceptibility, $\chi^d_{\sup}$, as a function of
           $\jt$, at densities $\langle n \rangle
           = 0.25$ ($\triangle$), $\quarter$ ($\square$),
           and $\langle n \rangle = 0.75$ ($\square$).

\item{1b}  Pairing-pairing correlation function
           ${\rm C({\bf m})}$
           as a function of
           distance ${\rm {\bf m}}$, at density
           $\quarter$ and $\jt = 3.0$. $\square$ denotes
           $\dx2y2$ pairing correlations, $\triangle$ indicates
           extended $s$ correlations, while $\square$ corresponds to
           ${\rm d_{xy}}$ correlations.

\item{1c}  Pairing-pairing correlation function
           ${\rm C( {\bf m} ) }$ in the $\dx2y2$ channel,
           as a function of
           distance ${\rm {\bf m}}$, at density
           $\quarter$. $\triangle$, $\square$ and $\square$ are results for
           $\jt = 1.0$, $3.0$ and $4.0$, respectively.

\item{2a}  Superfluid density, $\ds$, versus $\jt$, at several fermionic
densities.
           $\square$ corresponds to $\quarter$, $\triangle$ denotes
           results for $\langle n \rangle = 0.25$, while $\square$
           indicates $\langle n \rangle = 0.75$.

\item{2b}  Drude peak, $\drude$, as a function of $\jt$ for several
           densities. The notation is as in Fig.2a.

\item{3a}  Energy of the ground state
           as a function of an external magnetic flux $\phi$. The
           energy at zero flux is subtracted from the result i.e.
           $\Delta E ( \phi) = E(\phi) - E(0)$.
           The subspace of zero momentum is considered, and the
           density is $\quarter$. $\square$ denotes
           results at $\jt = 3.0$, while $\square$ corresponds to
           $\jt = 4.0$ i.e. inside the phase separated region.

\item{3b}  Spin-spin correlation ${\rm S({\bf m})}$ as a function of distance
at
           density $\quarter$, and coupling $\jt =  3.0$ i.e. in the
           superconducting region.

\item{3c}  Hole-hole correlations ${\rm h({\bf m})}$ as a function of distance
at
           density $\quarter$, and coupling $\jt =  3.0$ i.e. in the
           superconducting region.

\item{4a}  Qualitative representation of a ``dimer liquid'' state,
           presumed to be the ground state in the superconducting region
           discussed in this paper.

\item{4b}  Schematic $semi-quantitative$ phase diagram of the
           ${\rm t-J}$ model in two dimensions at zero temperature,
           as a function of coupling $\jt$, and hole density
           ${\rm x = 1 - \langle n \rangle}$.
           The curves separating the region at small $\jt$, presumably
           a Fermi liquid, FL, from
           the ``binding'' region, as well as the separation between
           binding and $\dx2y2$ wave superconductivity, are rough
           estimations
           based on the study of binding energies, and the strength of
           $\chid$.
           The transition leading to phase separation is more
           accurate, and in qualitative agreement with high temperature
           expansions.\refto{putikka}
           Near half-filling, the calculations are more difficult,
           and it is only known that antiferromagnetic,
           AF, and ferromagnetic, FM, correlations are important.

\endit